\documentclass{cpbtex}
\pdfoutput=1
\usepackage{graphicx}
\usepackage{subfigure}

\begin{document}

\title{Photon antibunching in a cavity-QED system with two Rydberg-Rydberg interaction atoms\thanks{Project supported by the National Natural Science Foundation of China (Grant No.11874190).}}

\author{Tong Huang$^{1}$, \ Lei Tan$^{1,2}$ \thanks{Corresponding author. E-mail:tanlei@lzu.edu.cn}\\
$^{1}${Lanzhou Center for Theoretical Physics, Key Laboratory of Theoretical Physics} \\
{of Gansu Province,Lanzhou University, Lanzhou, Gansu 730000, China}\\  
$^{2}${Key Laboratory for Magnetism and Magnetic Materials of the Ministry of Education}, \\
{Lanzhou University, Lanzhou $730000$, China}\\ 
}   


\date{\today}
\maketitle

\begin{abstract}
We propose how to achieve  strong photon antibunching  effect in a cavity-QED system coupled with two  Rydberg-Rydberg interaction atoms.
Via calculating the equal time second order correlation function $g^{(2)}(0)$, we find that  the unconventional photon blockade  and the conventional photon blockade appear in the atom-driven scheme,  and they are both significantly affected by the Rydberg-Rydberg interaction.
We also find that under appropriate parameters, the photon antibunching and  the mean  photon number can be significantly enhanced  by combining the conventional photon blockade and the unconventional photon blockade.
In the cavity-driven scheme, the existence of the Rydberg-Rydberg interaction severely destroys the  photon antibunching under the unconventional photon blockade mechanism. These results will  help to guide the implementation of the single photon emitter in the Rydberg atoms-cavity system.
\end{abstract}



\section{Introduction}
Photon antibunching is a quantum mechanic effect in which emitted photons tend to be detected one by one \ucite{1982Photon-antibunching,1977photon-antibunching,199Photon-antibunching,1986Generating,1986Bunching}. This important effect can be used to implement single photon emitter, which has important applications in quantum information processing \ucite{2004Quantum-dot,2007Linear}  and quantum computing \ucite{2005Single,2010On-chip}.
There are two methods to achieve strong photon antibunching  in the cavity-QED system.
One is called the conventional photon blockade (PB): A strongly coupled system between an optical cavity and another nonlinear degree of freedom possesses an anharmonic ladder in the energy spectrum. Thus, it can produce the photon  antibunching  when the single-photon nonlinearity is larger than  the mode linewidth \ucite{1994Possibility}.  In the past few decades, a sequence of experimental and theoretical works related to the PB have been investigated in various systems, such as cavity-atom systems \ucite{2002Deterministic,2016Fate}, superconducting qubit systems \ucite{2014From,2011Ph}, optomechanical systems \ucite{2013Photon-induced,2013Photon-blockade,2011Phot}, etc.

In 2010, Liew and Savona  found another mechanism called the unconventional photon blockade (UPB)  originated from  quantum interference which leads  the probability amplitude  of the two photon state  to zero \ucite{2010Single}. The UPB usually requires  to add  additional degrees of freedom, such as auxiliary cavities \ucite{2017Unconventional,2015Unconventional,2011Origin} or two-level atoms (TLAs) \ucite{2019A-Photon,2019Antibunching} into the system in order to establish multiple transition pathways to realize the destructive interference.
In contrast with the PB, this interference-based photon  blockade mechanism can achieve strong photon antibunching only requiring weak nonlinearity \ucite{2017Unconventional,2018Observation1,2018Observation2,2000Intensity}.
Due to  this significant  feature, a large number of  works related to  the UPB have been reported extensively in  various system, such as single mode cavity with second order nonlinearity \ucite{2014Unconventional-photon}, single mode cavity with  Kerr-type nonlinearity \ucite{2010Single,2013Optimal,2014Tun}, two-level emitter-cavity system \ucite{2019Antibunching}, single mode cavity including a degenerate optical parametric amplifier \ucite{2017Quantum-interference-assisted}, coupled optomechanical system  \ucite{2015Tunable,2020Photon} and Gaussian squeezed states \ucite{2014Antibunching}.

Rydberg atoms  with  large principal quantum numbers $n$ have become an important tool of quantum information processing \ucite{2010Quantuminformation,2016Quantumcomputing}.  Rydberg blockade  appears in the Rydberg atoms ensemble because of the strong Rydberg-Rydberg interaction, where  Rydberg excitation of one atom prevents the excitation of atoms within the blockade radius. This phenomenon can be used to induce strong optical nonlinearity \ucite{2016Nonlinear} which generates  nonclassical states of light  in the Rydberg atoms ensemble and the Rydberg atoms-cavity system  \ucite{2013PhotonRydberg,2014Quantumstatistics,2012CorrelatedPhoton}. In addition,  photon antibunching induced  by quantum interference can also be obtained in  TLAs-cavity system\ucite{2019Interfering,2019Hybrid}.  Thus, one expects to  obtain strong photon antibunching  in two level Rydberg atoms-cavity system by combining the above two effects.

This  paper is organized as follows. In Sec. \uppercase\expandafter{\romannumeral2}, we illustrate the theoretical model and Hamiltonian of the system, the cavity-atom interaction part of the Hamiltonian is rewritten in terms of the collective operators to obtain an intuitive picture of the photon  transition pathways. In Sec. \uppercase\expandafter{\romannumeral3}, we calculate the equal time second order correlation function under the cavity-driven  and the atom-driven scheme, respectively. Numerical and analytical methods are used to obtain the effect of Rydberg-Rydberg interaction on the photon antibunching. In addition, a summary is given in  Sec. \uppercase\expandafter{\romannumeral4}.

\section{Model}
\begin{figure}[!htbp]
\centering  
\includegraphics[width=0.62\textwidth]{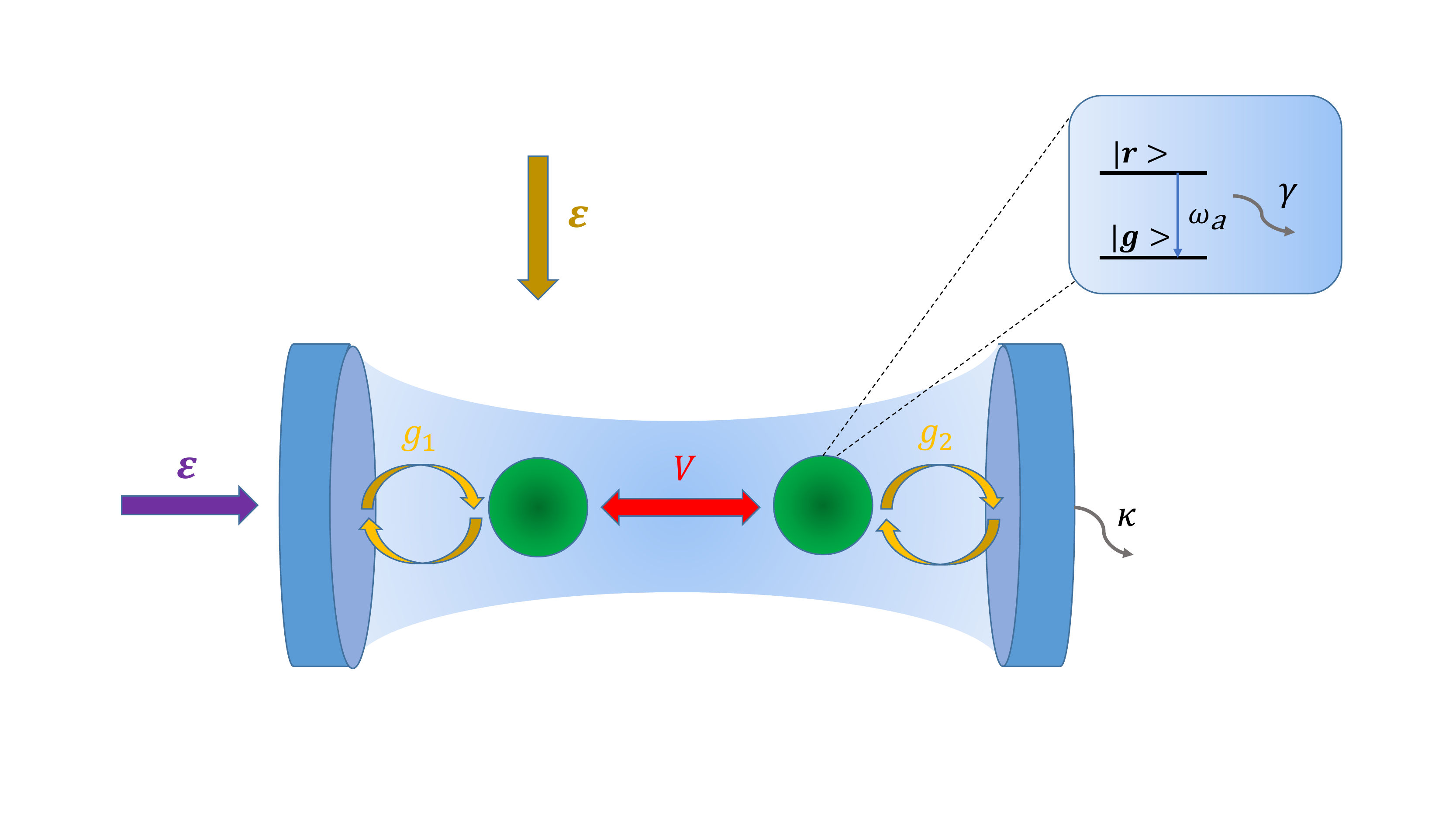}
\caption{Scheme of a dissipative system: two identical two level Rydberg atoms are  trapped in the single model cavity. The atoms  are pumped in the direction perpendicular to the axis of the cavity and the cavity is pumped in the direction parallel to the axis of the cavity.}
\label{fig1}
\end{figure}

As shown in Fig. \ref{fig1}, the system consists of a single mode cavity and two Rydberg coupling atoms  trapped alone the axis of the cavity. A continuous wave coherent field is applied to drive the atoms (cavity). The total Hamiltonian of the system can be written in the rotating frame  with  the laser frequency $\omega_{L}$ as
\begin{equation}
\hat{H}=-\sum_{j=1,2}\Delta_{a}\sigma_{rr}^{j} -\Delta_{c}a^{\dag}a  +\sum_{j=1,2}g_{j}(\sigma_{rg}^{j}a+\sigma_{gr}^{j}a^{\dag})+V\sigma_{rr}^{1}\sigma_{rr}^{2}+\hat{H}_{d}. \label{e1}
\end{equation}
where $a$ ($a^{\dag}$) is the annihilation (creation) operator of the  cavity mode with resonant frequency  $\omega_{c}$. $\sigma_{mn}^{j}=|m \rangle_{j}$$_{j} \langle n|$ with $m, n=r, g$ and  the transition frequency between the ground state $|g\rangle$ and the Rydberg state $|r\rangle$ is denoted by  $\omega_{a}$.
$g_{j}=gcos(kx_{j})$  is the position dependent coupling  strength between the cavity field  and the $jth$ atom at the point $x_{j}$ \ucite{2017Hyperradiance}. $V$ represents the Rydberg  coupling strength  between  two  Rydberg atoms  at a distance of $d$ , it takes the form $V\propto \frac{1}{d^{6}}$ when Rydberg atoms interact through a van der Waals potential and the form $V\propto \frac{1}{d^{3}}$ when Rydberg atoms couple through the dipole-dipole interaction . The  last term $\hat{H}_{d}$ in Eq. (\ref{e1}) describes  cavity-driven term with $\varepsilon(a+a^{\dag}) $ and  atom-driven term  with $\varepsilon\sum_{j=1,2}(\sigma_{rg}^{j}+\sigma_{gr}^{j})$, respectively. Here  $\Delta_{c}=\omega_{L}-\omega_{c}$ denotes the laser-cavity frequency detuning. $\Delta_{a}=\omega_{L}-\omega_{a}$ is the laser-atom frequency detuning and $\varepsilon$ represents the pump laser amplitude.

Defining the  collective operators \ucite{2017Collective} as
\begin{equation}
D_{\pm}^{\dag}=\frac{1}{\sqrt{2}}(\sigma_{rg}^{1}\pm \sigma_{rg}^{2}). \label{e3}
\end{equation}
and the atom-cavity interaction part  of the Hamiltonian in Eq. (\ref{e1}) can be reformulated as
\begin{equation}
\hat{H}_{I}=\hat{H}_{+}+\hat{H}_{-}. \label{e4}
\end{equation}
\begin{equation}
\hat{H}_{\pm}=\frac{g}{\sqrt{2}}(cos(kx_{1})\pm cos(kx_{2}))(aD_{\pm}^{\dag}+a^{\dag}D_{\pm}). \label{e5}
\end{equation}
where $\hat{H}_{+}$ represents the coupling of the symmetric state and the cavity mode  leading to the transitions $|gg,n+2\rangle\longleftrightarrow|+,n+1\rangle\longleftrightarrow|rr,n\rangle$.
$\hat{H}_{-}$ represents the coupling of the antisymmetric state and the cavity mode, which gives rise to the transitions
$|gg,n+2\rangle\longleftrightarrow|-,n+1\rangle\longleftrightarrow|rr,n\rangle$.
 If we choose the distance between two Rydberg aroms as $d=x_{2}-x_{1}=n\lambda$ ( $n=0,1,2...$), $\hat{H}_{-}$ vanishes and the cavity mode only couples via $|+\rangle$.  Thus, the Hamiltonian of the system  can be rewritten as
\begin{equation}
\hat{H}_{r}=-\sum_{j=1,2}\Delta_{a}\sigma_{rr}^{j} -\Delta_{c}a^{\dag}a  +g\sum_{j=1,2}(\sigma_{rg}^{j}a+\sigma_{gr}^{j}a^{\dag})+V\sigma_{rr}^{1}\sigma_{rr}^{2}+\hat{H}_{d}. \label{e6}
\end{equation}
In the following, we will discuss the photon antibunching of the system with atom-driven and cavity-driven  scheme, respectively.

\section{Results and discussions}

\subsection{atom-driven scheme}
\begin{figure}[!htbp]
\centering  
\includegraphics[width=0.62\textwidth]{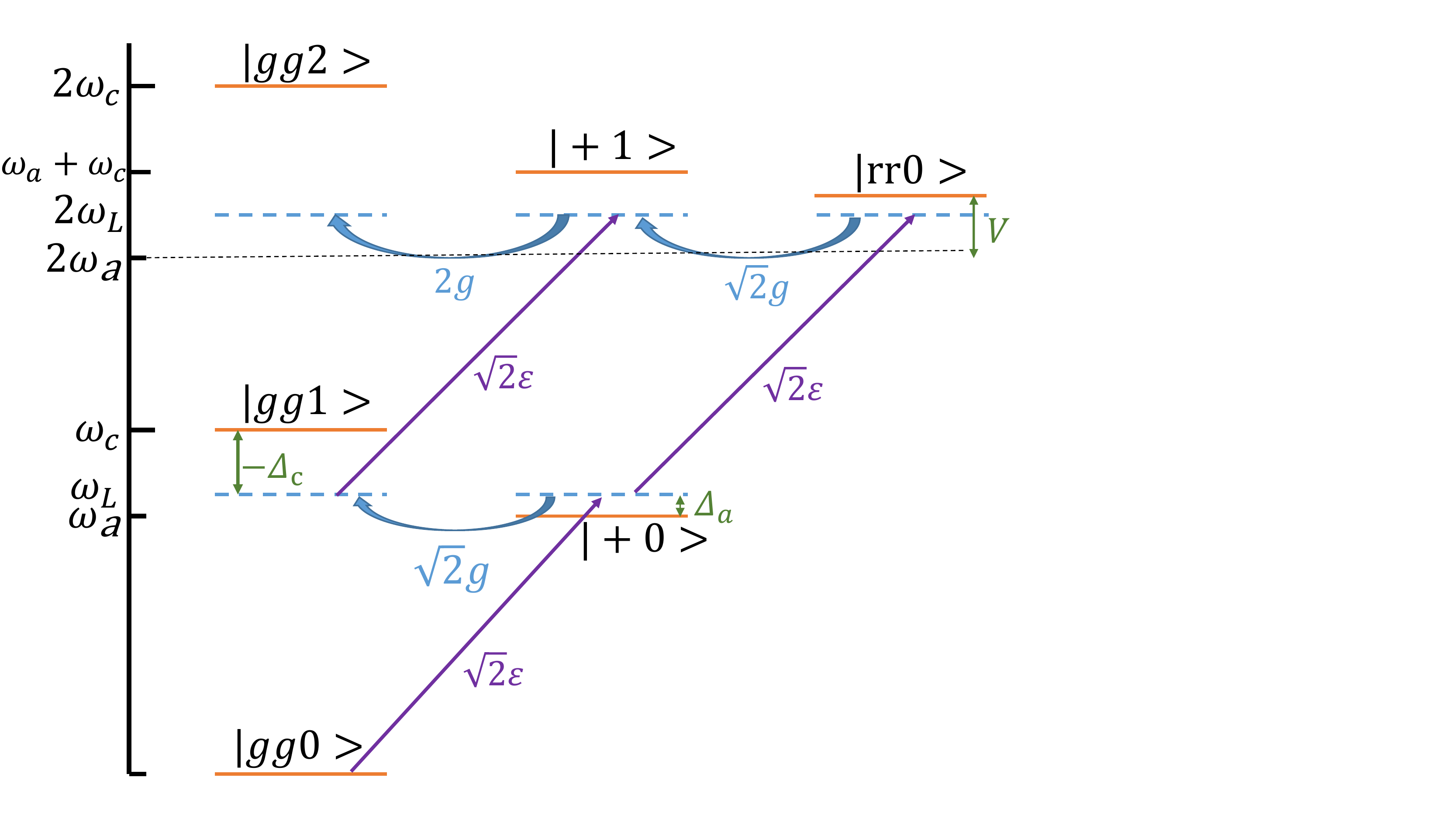}
\caption{ Transition pathways with atom-driven scheme }
\label{fig2}
\end{figure}

First, in order to get instructive understanding of the photon antibunching, one seeks to  analytical description of  the equal time  second  order correlation function $g^{(2)}(0)$, which measures the  photon statistic distribution of the cavity mode.
Within the  weak  driving  condition, only lower photons excitation states are occupied, then one  can express the wave function of  the system with  truncated state as follows
\begin{eqnarray}
|\psi(t\rangle)=C_{gg0}(t)|gg0\rangle+C_{gg1}(t)|gg1\rangle+C_{gg2}(t)|gg2\rangle+\nonumber\\
C_{+0}(t)|+0\rangle+C_{+1}(t)|+1\rangle+C_{rr0}(t)|rr\rangle. \label{e7}
\end{eqnarray}
the coefficient $C_{aan}(t)$ represents the probability amplitude of the  state $|aan\rangle=|aa\rangle\otimes |n\rangle$, where $|aa\rangle (aa=gg,rr,+)$ is the collective states of the two coupled Rydberg atoms and $|n\rangle (n=0,1,2)$ is the Fock basis for the cavity mode. The value $|C_{aan}(t)|^{2}$  denotes  the  probability of the  corresponding state. In order to obtain the value of the probability amplitude, we need to solve the time-dependent Schr\"{o}dinger equation $i\hbar \frac{\partial}{\partial_{t}}|\psi(t)\rangle=\widetilde{H}|\psi(t)\rangle$, where $\widetilde{H}$ is the effective non-Hermitian Hamiltonian

\begin{equation}
\widetilde{H}=\hat{H}_{r}-\frac{i\kappa a^{\dag}a}{2}-\frac{i\gamma }{2}(\sigma_{1}^{\dag}\sigma_{1}+\sigma_{2}^{\dag}\sigma_{2}). \label{e8}
\end{equation}
$\kappa$ and $ \gamma$ are  loss rates of the cavity and the atoms, respectively.
Substituting  Eq. (\ref{e7}) and  Eq. (\ref{e8}) into the Schr\"{o}dinger equation, one can obtain a set of equations of the time-dependent coefficients ($\hbar=1$)
\begin{equation}
\begin{split}
&i\dot{C}_{gg1}=\sqrt{2}g C_{+0}+\sqrt{2}\varepsilon C_{+1}-\Delta_{c}C_{gg1}-\frac{i\kappa}{2}C_{gg1},\\
&i\dot{C}_{gg2}=2gC_{+1}-2\Delta_{c}C_{gg2}-i\kappa C_{gg2},\\
&i\dot{C}_{rr0}=(-2\Delta_{a}+V-i\gamma)C_{rr0}+\sqrt{2}gC_{+1}+\sqrt{2}\varepsilon C_{+0},\\
&i\dot{C}_{+0}=-\Delta_{a}C_{+0}+\sqrt{2}g C_{gg1}+\sqrt{2}\varepsilon (C_{gg0}+C_{rr0})-\frac{i\gamma}{2}C_{+0},\\
&i\dot{C}_{+1}=-(\Delta_{a}+\Delta_{c})C_{+1}+\sqrt{2}gC_{rr0}+2gC_{gg2}+\sqrt{2}\varepsilon C_{gg1}-\frac{i(\gamma+\kappa)}{2}C_{+1}. \label{e9}
\end{split}
\end{equation}

Under the weak driving condition, given that $\{C_{gg0}\}\gg \{C_{gg1}, C_{+0}\}\gg\{ C_{gg2}, C_{+1}, C_{rr0}\}$ and set $C_{gg0}=1$,
the steady state  solution  can  be  obtained as follows
\begin{equation}
\begin{split}
&C_{gg1}=-\frac{8g\varepsilon}{M},\\
&C_{gg2}=\frac{16\sqrt{2}g^{2}\varepsilon^{2}[\kappa+2\gamma-4i\Delta_{a}-2i\Delta_{c}+2iV)]}{MN}. \label{e10}
\end{split}
\end{equation}
where $M=8g^2-2i\kappa\Delta_{a}-2i\gamma\Delta_{c}-4\Delta_{a}\Delta_{c}+\kappa\gamma$, $N=4g^2\kappa-16ig^2\Delta_{a}+2i\kappa^{2}\Delta_{a}-4\kappa\Delta_{a}^{2}-8ig^{2}\Delta_{c}+8i\Delta_{a}^{2}\Delta_{c}+8i\Delta_{a}\Delta_{c}^{2}+8g^2\gamma-\kappa^{2}\gamma-4\Delta_{c}^{2}\gamma-\kappa\gamma^{2}+2i\Delta_{c}\gamma^{2}+8ig^{2}V-i\kappa^{2}V+2\kappa\Delta_{a}V-4i\Delta_{a}\Delta_{c}V-4i\Delta_{c}^{2}V-i\kappa\gamma V+2i\Delta_{c}\gamma V$.
Under the  condition of the large laser-atom (laser-cavity) frequency detuning limit: $\Delta_{a}, \Delta_{c} \gg \gamma, \kappa $, an analytical expression of the equal time second order correlation function can be represented as
\begin{equation}
g^{(2)}(0)=\frac{<a^{\dag}a^{\dag}aa>}{<a^{\dag}a>^{2}}\simeq \frac{2|C_{gg2}|^{2}}{|C_{gg1}|^4}
\simeq -\frac{16(2g^{2}-\Delta_{a}\Delta_{c})^2(2\Delta_{a}+\Delta_{c}-V)^2}{N^{2}}.
\end{equation}
which quantifies the joint  probability  of detecting two photons at the same time.
In the limitation  of   $g^2(0)\rightarrow 0$, one can obtain the optimal conditions of the photon antibunching
\begin{equation}
\begin{split}
&\Delta_{a}=\frac{1}{2}(V-\Delta_{c}),\\
&\Delta_{a}=\frac{2g^{2}}{\Delta_{c}}.
\end{split} \label{e12}
\end{equation}

The first expression in Eq. (\ref{e12}) is an  optimal  condition for the UPB  induced by  quantum interference, which can be obtained when  the probability amplitude  $C_{gg2}$=0.  One can see that  it's relevant to  the laser-atom  frequency  detuning, laser-cavity  frequency detuning,  and Rydberg coupling strength. As shown in Fig. \ref{fig2}, two transition pathways  $|gg0\rangle\stackrel{\sqrt{2}\varepsilon}{\longrightarrow}|+0\rangle\stackrel{\sqrt{2}g}{\longrightarrow}|gg1\rangle
\stackrel{\sqrt{2}\varepsilon}{\longrightarrow}|+1\rangle\stackrel{2g}{\longrightarrow}|gg2\rangle$
and $|gg0\rangle\stackrel{\sqrt{2}\varepsilon}{\longrightarrow}|+0\rangle\stackrel{\sqrt{2}\varepsilon}{\longrightarrow}|rr0\rangle
\stackrel{\sqrt{2}g}{\longrightarrow}|+1\rangle\stackrel{2g}{\longrightarrow}|gg2\rangle$ are distinguishable. Thus, destructive interference arises between the two pathways and one can  expect to get a strong UPB effect \ucite{2017Hyperradiance}. The second formula in Eq. (\ref{e12}) is the  optimal  condition of the PB. As shown in previous studies \ucite{2019Interfering}, one can obtain a PB effect when the cavity-atom coupling strength $g$ is  large  enough.

Next, we will numerically calculate the equal time second order correlation function $g^{(2)}(0)$ by solving the  master equation of the system

\begin{equation}
i\frac{\partial \hat{\rho}}{\partial t}=[\hat{H} , \hat{\rho}]-i\sum_{j=1,2}\frac{\gamma_{j}}{2} \mathcal{D}[\sigma_{j}]\hat{\rho}  -i\frac{\kappa}{2} \mathcal{D}[a]\hat{\rho}. \label{e13}
\end{equation}
where $\mathcal{D}[q]\hat{\rho}=2q \rho q^{\dag}-q^{\dag}q\hat{\rho}-\hat{\rho}q^{\dag}q$ denotes the  Liouvillian   superoperator that represents the loss of the Rydberg atoms (cavity), the collective damping of the Rydberg atoms is not considered  for simplicity.  The  steady state  solution of the equal time second order correlation function  $g^{(2)}(0)$  is  expressed as
\begin{equation}
g^{(2)}(0)=\frac{Tr[\hat{\rho} a^{\dag}a^{\dag}aa]}{(Tr[\hat{\rho} a^{\dag}a])^2}. \label{e14}
\end{equation}

 In the experiment, we consider two $^{87}Rb$ atoms placed in a plane cavity to produce nonclassical light. Such a single photon emitter can be realized by implementing a two-dimensional optical lattice in a cavity \ucite{2019Deterministic}, and the two-dimensional optical lattice consists of  a red-detuned laser beam  perpendicular to the cavity axis and a blue-detuned laser beam which alone the cavity axis. Rydberg atoms are distributed at several lattice sites, so they have well-defined positions. Moreover, excess Rydberg atoms  in  the optical lattice  can be removed with a resonant push-out beam.
 The experimentally achievable parameters of the  distance between two $^{87}Rb$ atoms can be adjusted from 15$\mu$m to 4$\mu$m \ucite{2013Direct}. For the  Rydberg states $|r\rangle$=$|62D_{\frac{3}{2}}\rangle$, the van der Waals interaction cofficient is $C_{6}=730GHz.\mu$$m^{6}$, and the Rydberg coupling strength $V$ can be tuned continuously between $2\pi\times(0.01,28.33)MHz$ \ucite{2013Direct}. In addition, for principal quantum number $n\sim$ 60,  the $^{87}Rb$ atoms $|nD_{\frac{3}{2}}\rangle$ state has a lifetime of 203.80$\mu$s (the spontaneous decay rate is $\gamma=2\pi\times0.4KHz$ \ucite{2009Quasi, 2021Efficient}).  The pump laser amplitude $\varepsilon$ can be adjusted from $2\pi\times500KHz$ to $2\pi\times5MHz$. Moreover, the cavity QED experiments with $^{87}Rb$ atoms provide us other parameters: $(g,\kappa)=2\pi\times(7.8,2.5)MHz$ \ucite{2019Deterministic}. Next, we  numerically calculate the photon antibunching by using the above experimental parameters.

\begin{figure}[!htbp]
\centering  
\subfigure[]{
\label{fig3.1}
\includegraphics[width=0.48\textwidth]{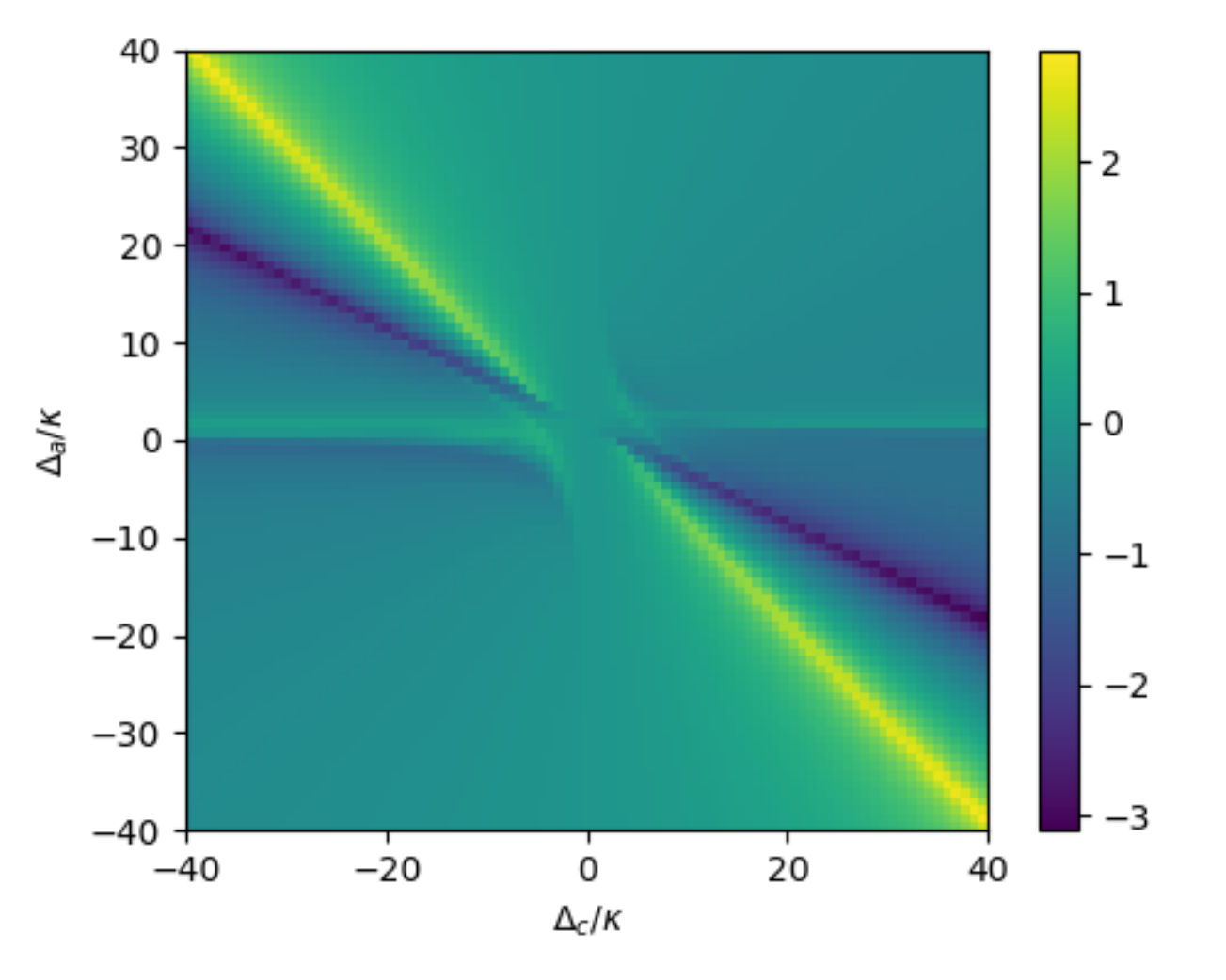}}
\subfigure[]{
\label{fig3.2}
\includegraphics[width=0.48\textwidth]{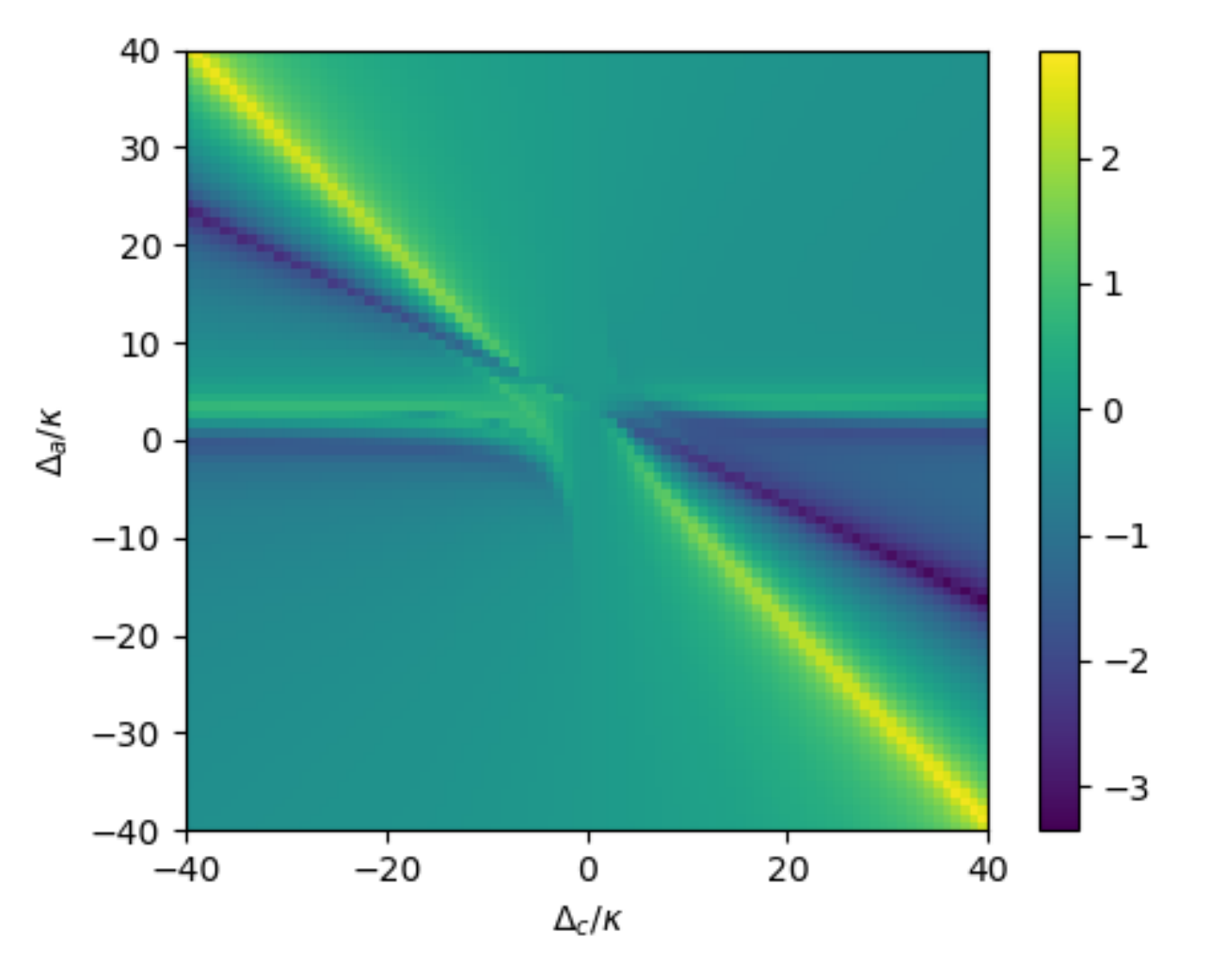}}
\caption{(a) Plot of $log_{10}[g^{(2)}(0)]$ as a function of $\Delta_{c}/\kappa$ and $\Delta_{a}/\kappa$ with   $V=2\kappa$.
(b) Plot of $log_{10}[g^{(2)}(0)]$ as a function of $\Delta_{c}/\kappa$ and $\Delta_{a}/\kappa$ with $V=6\kappa$.  Other parameters take as $\gamma=2\pi\times0.4KHz$, $\kappa=2\pi\times2.5MHz$, $\varepsilon= 0.4\kappa$ and $g=2\pi\times7.8MHz$.
}
\label{fig3}
\end{figure}

In  order to show the dependence of the $g^{(2)}(0)$ on the Rydberg coupling strength $V$, one  plots  the equal time second order correlation function $g^{(2)}(0)$   (in logarithmic units)  as a function  of  $\Delta_{a}/\kappa$ and $\Delta_{c}/\kappa$ with the   Rydberg coupling strength $V=2\kappa$ and 6$\kappa$. The atom-driven strength $\varepsilon$ is  assigned to 0.4$\kappa$  for weak driving condition.
From  Figs. \ref{fig3.1} and  \ref{fig3.2}, one can  see  that the   UPB effect  appears   and the position of the UPB can be changed  by the  Rydberg coupling strength $V$. We also find that the PB effect becomes significant in the case of the large cavity-atom coupling strength $g$.
These numerical results are consistent with the analytical solution for the optimal conditions in Eq. (\ref{e12}).

\begin{figure}[!htbp]
\centering  
\subfigure[]{
\label{fig4.1}
\includegraphics[width=0.48\textwidth]{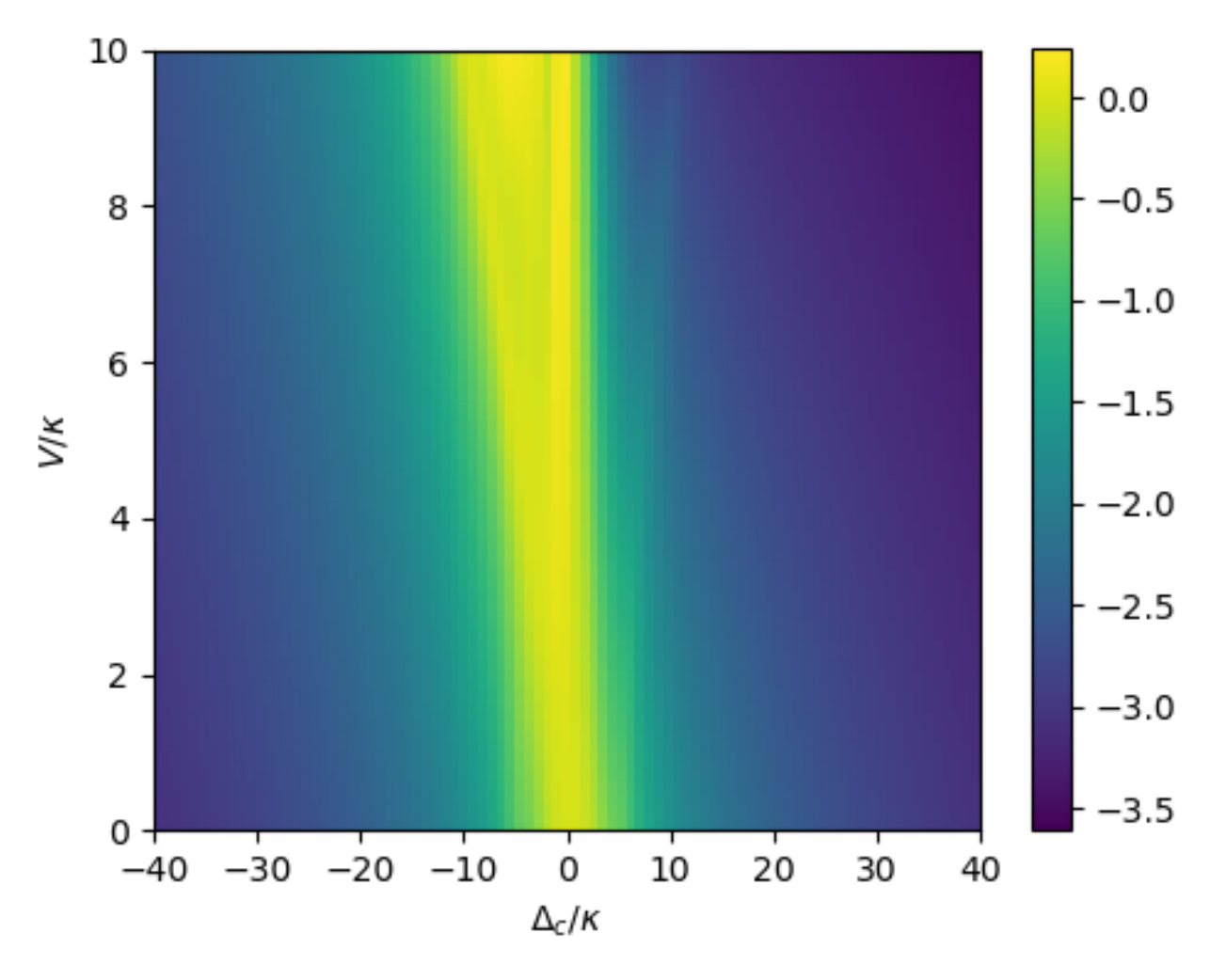}}
\subfigure[]{
\label{fig4.2}
\includegraphics[width=0.48\textwidth]{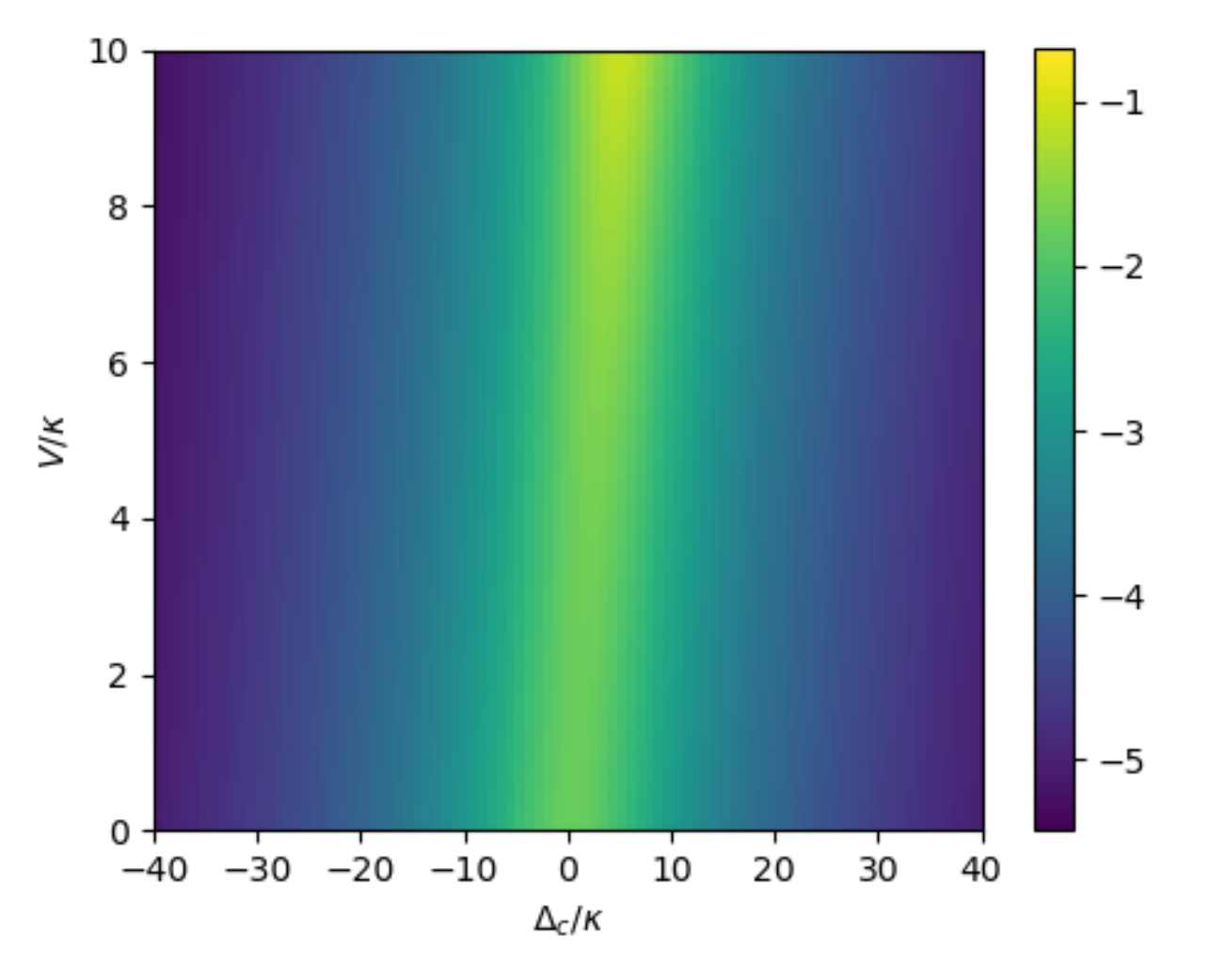}}
\caption{(a) Plot of $log_{10}[g^{(2)}(0)]$ as a function of $\Delta_{c}/\kappa$  and  $V/\kappa$ with $\Delta_{a}=-\frac{1}{2}(\Delta_{c}-V)$.
(b) Plot of the mean photon number $\langle a^{\dag}a \rangle$  (log scale)  as a function of $\Delta_{c}/\kappa$  and  $V/\kappa$ with $\Delta_{a}=-\frac{1}{2}(\Delta_{c}-V)$.  Other parameters take as $\gamma=2\pi\times0.4KHz$, $\kappa=2\pi\times2.5MHz$, $\varepsilon= 0.4\kappa$ and $g=2\pi\times7.8MHz$.
}
\label{fig4}
\end{figure}

To further clarify the effect of the Rydberg coupling  strength $V$ on  the UPB, we  plot the equal time second order correlation function  $g^{(2)}(0)$ and the mean photon number $\langle a^{\dag}a \rangle$   as a function of $\Delta_{c}/\kappa$  and  $V/\kappa$ with $\Delta_{a}=-\frac{1}{2}(\Delta_{c}-V)$ in Figs. \ref{fig4.1} and \ref{fig4.2}. From  Fig. \ref{fig4.1},  one can see that photon antibunching based on the quantum interference  is greatly enhanced by the laser-cavity frequency detuning $\Delta_{c}$.
Under the  condition of the large negative laser-cavity frequency detuning limit,  the smaller the  Rydberg coupling strength $V$  the stronger the UPB effect is. In the limit of the large positive laser-cavity frequency detuning, there is also a strong UPB effect at any value of $V$.
Contrary to the second-order correlation function $g^{(2)}(0)$, the mean photon number $\langle a^{\dag}a \rangle$ in the cavity  is  suppressed by the laser-cavity frequency detuning.
This brings difficulties to the detection of single photons under strong photon antibunching conditions caused by quantum interference.

\begin{figure}[!htbp]
\centering  
\subfigure[]{
\label{fig5.1}
\includegraphics[width=0.48\textwidth]{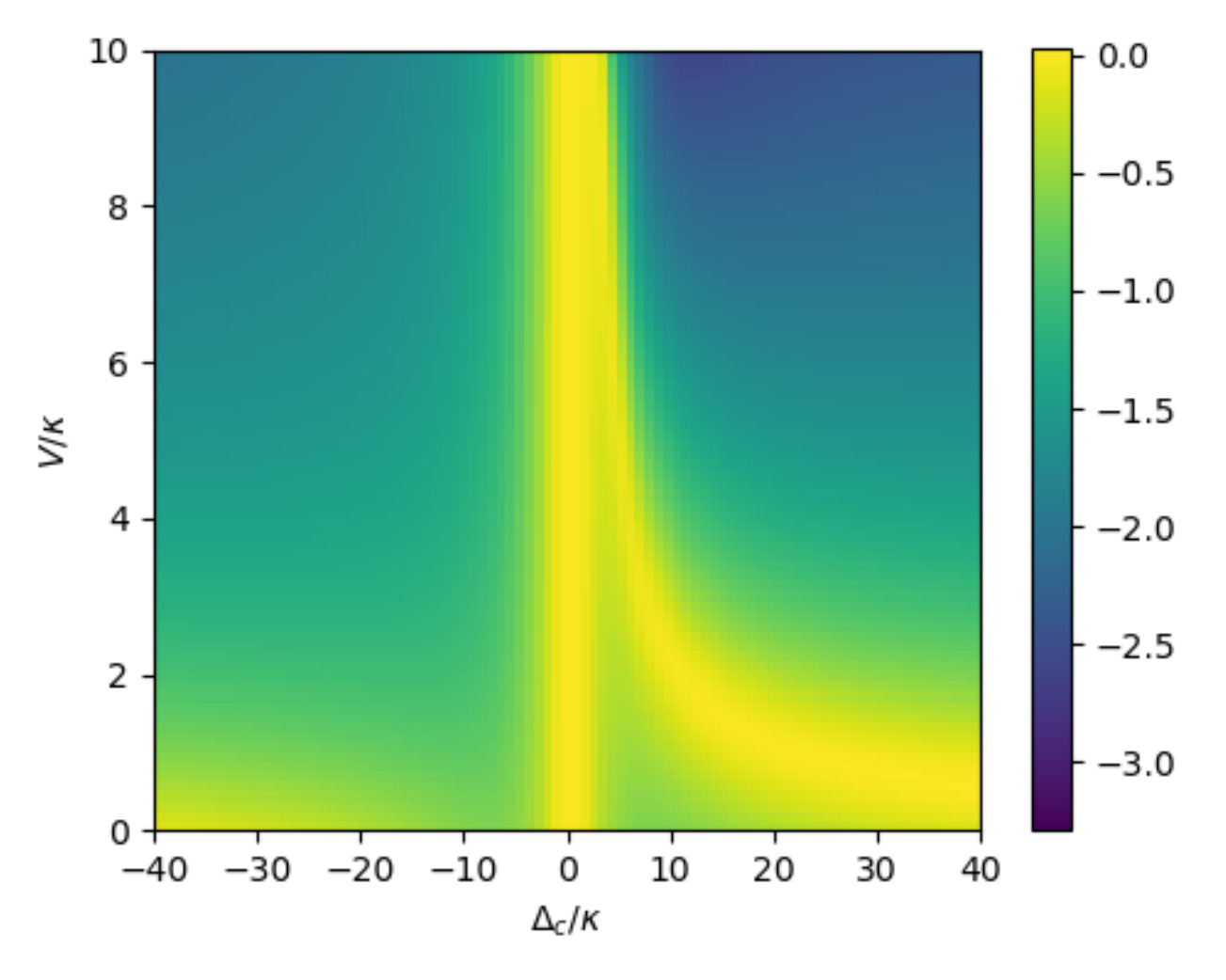}}
\subfigure[]{
\label{fig5.2}
\includegraphics[width=0.48\textwidth]{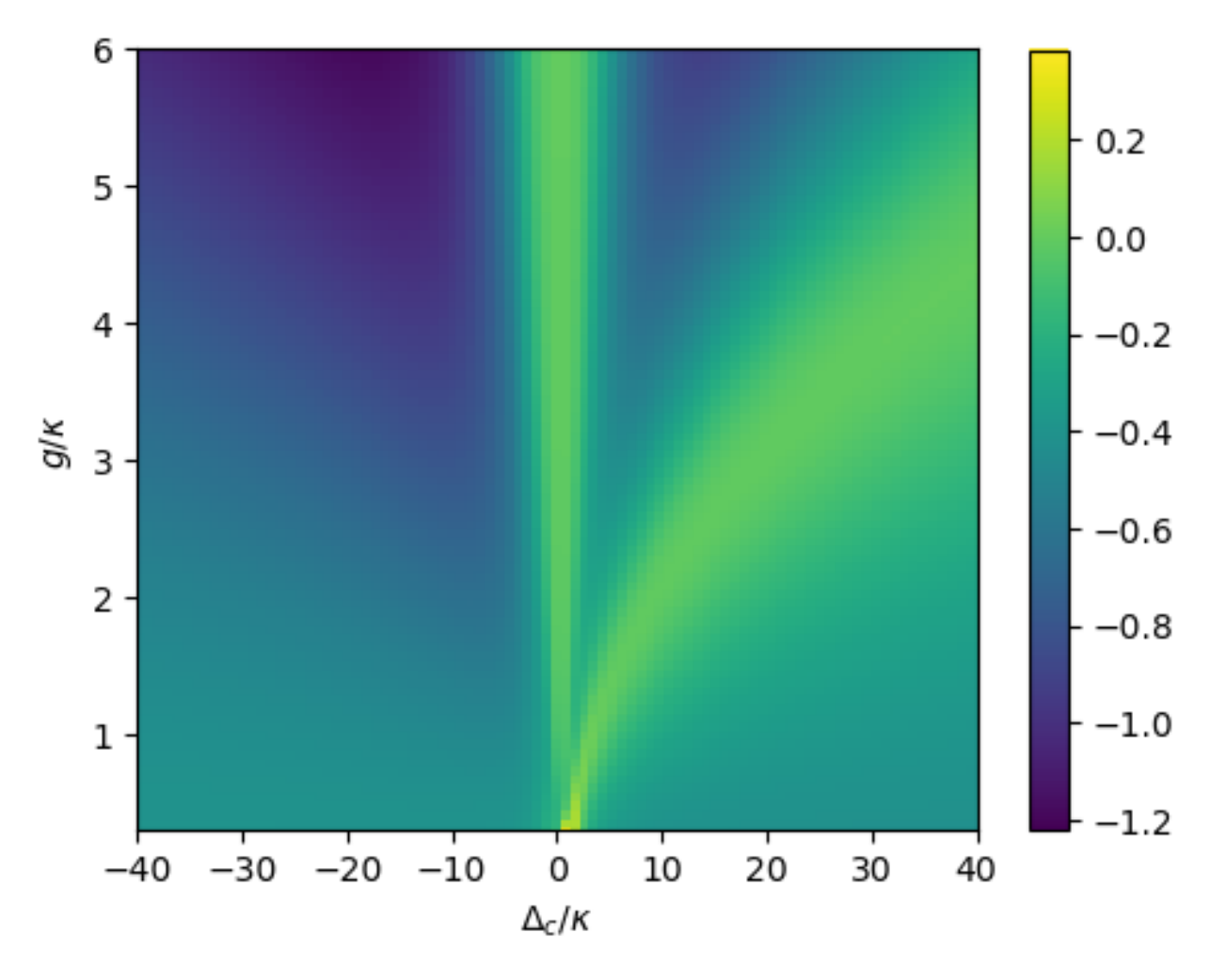}}
\caption{(a) Plot of $log_{10}[g^{(2)}(0)]$ as a function of $\Delta_{c}/\kappa$ and $V/\kappa$ with  $\Delta_{a}=\frac{2g^{2}}{\Delta_{c}}$, the atom-cavity coupling strength is given by $g=2\pi\times7.8MHz$.
(b) Plot of $log_{10}[g^{(2)}(0)]$ as a function of $\Delta_{c}/\kappa$ and $g/\kappa$ with $\Delta_{a}=\frac{2g^{2}}{\Delta_{c}}$, the Rydberg coupling strength is given by $V=\kappa$.  Other parameters take as $\gamma=2\pi\times0.4KHz$, $\kappa=2\pi\times2.5MHz$, $\varepsilon= 0.4\kappa$.
}
\label{fig5}
\end{figure}

Next, we investigate the effect of the Rydberg coupling strength $V$ and the cavity-atom coupling strength  $g$ on the PB under the optimal condition $\Delta_{a}=\frac{2g^{2}}{\Delta_{c}}$.  In  Fig. \ref{fig5.1},  we  plot the equal time second order correlation function  $g^{(2)}(0)$   as a function of $\Delta_{c}/\kappa$  and  $V/\kappa$. It's clear that  strong  photon antibunching based on the energy-level anharmonicity  can be obtained in the positive laser-cavity frequency detuning interval  when the Rydberg coupling strength $V$ is large enough.
Fig. \ref{fig5.2} displays the equal time second order correlation function $g^{(2)}(0)$ as a function of $\Delta_{c}/\kappa$  and  $g/\kappa$.
We find that under the PB optimal condition, the PB effect can be enhanced with the increase of coupling strength $g$ at the large laser-cavity frequency detuning limit.

\begin{figure}[!htbp]
\centering  
\subfigure[]{
\label{fig6.1}
\includegraphics[width=0.48\textwidth]{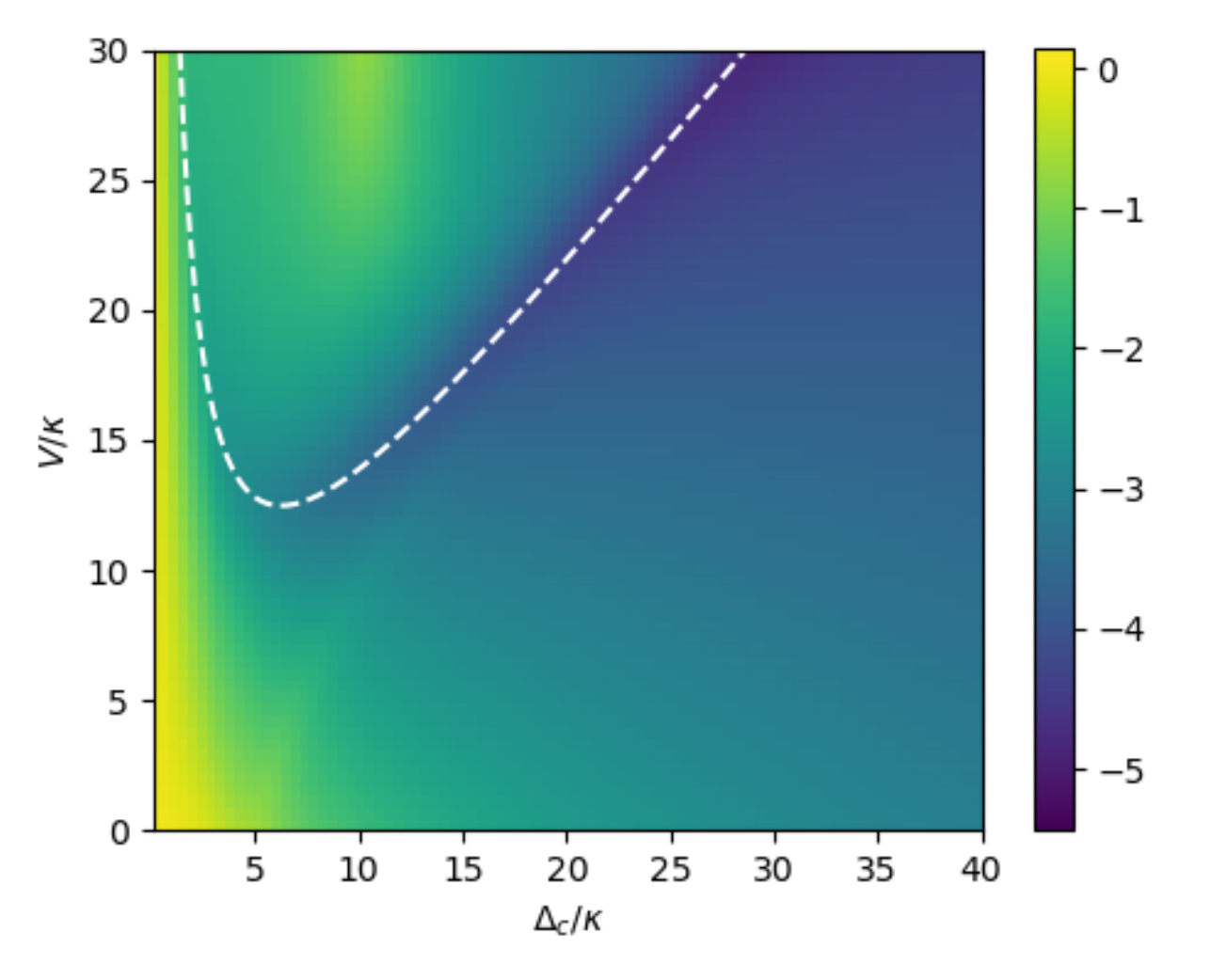}}
\subfigure[]{
\label{fig6.2}
\includegraphics[width=0.48\textwidth]{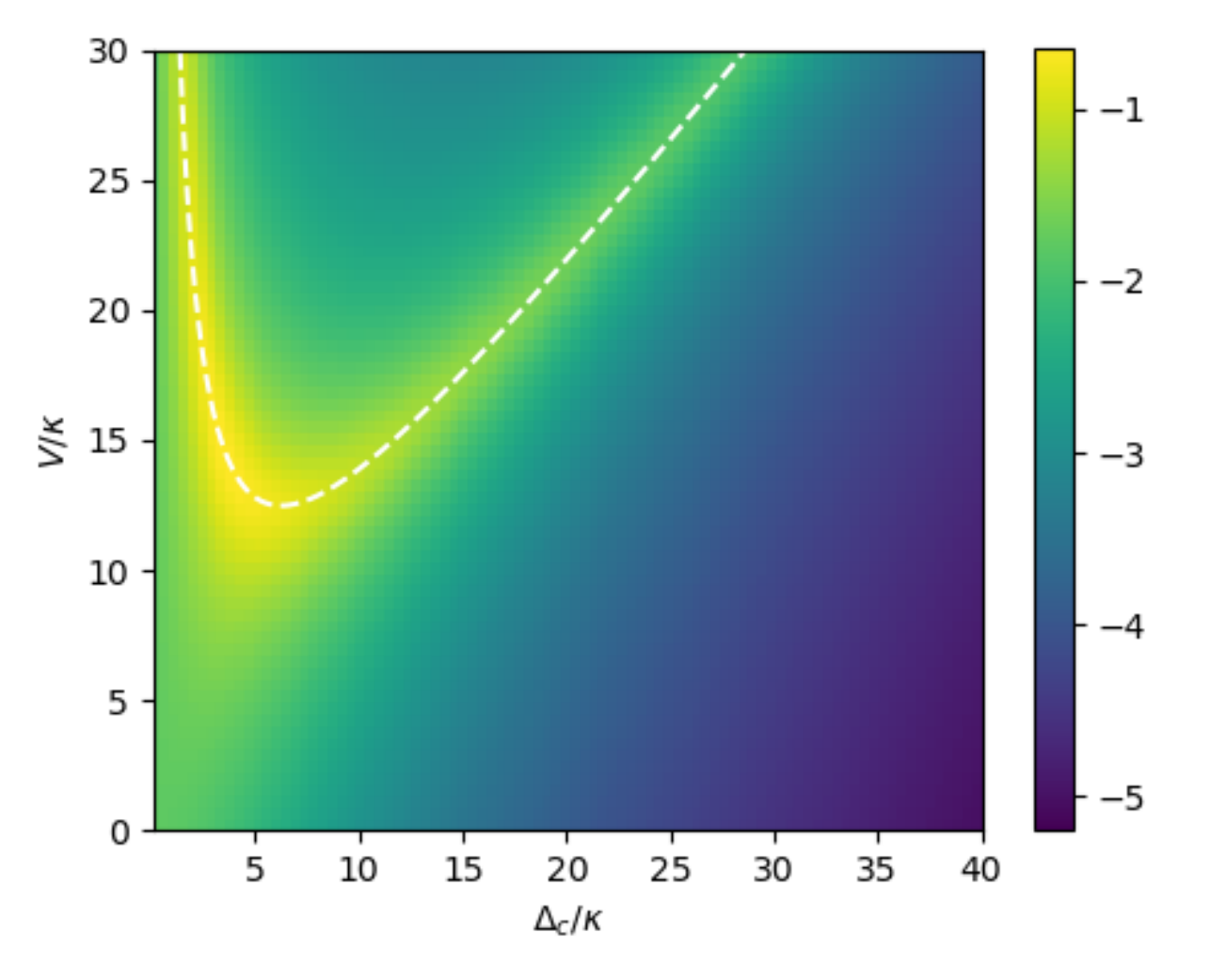}}
\caption{(a) Plot of $log_{10}[g^{(2)}(0)]$ as a function of $\Delta_{c}/\kappa$ and $V/\kappa$ with $\Delta_{a}=-\frac{1}{2}(\Delta_{c}-V)$.
(b) Plot of  $\langle a^{\dag}a \rangle$ (log scale) as a function of $\Delta_{c}/\kappa$ and $V/\kappa$ with $\Delta_{a}=-\frac{1}{2}(\Delta_{c}-V)$.  Other parameters take as $\gamma=2\pi\times0.4KHz$, $\kappa=2\pi\times2.5MHz$, $\varepsilon= 0.4\kappa$ and $g=2\pi\times7.8MHz$. The white dashed  line is depicted  by the  function $V=\Delta_{c}+\frac{4g^{2}}{\Delta_{c}}$.
}
\label{fig6}
\end{figure}

The above analysis in Fig. \ref{fig4} and Fig. \ref{fig5} are based on the separate consideration of the UPB  and the PB effect. Thus, we can only get a small mean photon number $\langle a^{\dag}a \rangle$  when the equal time second order correlation function $g^{(2)}(0)$ is extremely small. It's an adverse result for single photons detection. Here, we consider whether we can obtain stronger photon antibunching  and  larger mean photon number in the cavity by combining the  UPB effect and  the PB effect.
In Figs. \ref{fig6.1} and  \ref{fig6.2}, we  plot the equal time second order correlation function  $g^{(2)}(0)$ and the mean photon number $\langle a^{\dag}a \rangle$  as a function of $\Delta_{c}/\kappa$  and  $V/\kappa$ with $\Delta_{a}=-\frac{1}{2}(\Delta_{c}-V)$.
As shown in Fig. \ref{fig6.1}, the value of the equal time second order correlation function $g^{(2)}(0)$  at the  minimum is about $10^{-5.5}$.
As assumed above, a stronger photon antibunching based on the UPB and the PB mechanism appears when the parameters satisfy  $V \textgreater  4g$. The white dashed line $(V=\Delta_{c}+\frac{4g^{2}}{\Delta_{c}})$ in  Fig. \ref{fig6.1} corresponds to the intersection of the UPB optimal condition and the PB optimal condition. It analytically marks the minimum value of the equal time second order correlation function $g^{(2)}(0)$ in the parameter space.
In Fig. \ref{fig6.2},  we can see  that under the condition that the parameters  satisfy the UPB optimal condition, the mean photon number $\langle a^{\dag}a \rangle$ in the cavity can reach the maximum simultaneously when the equal time second order correlation function $g^{(2)}(0)$ takes a minimum value.

\begin{figure}[!htbp]
\centering  
\subfigure[]{
\label{fig9.1}
\includegraphics[width=0.41\textwidth]{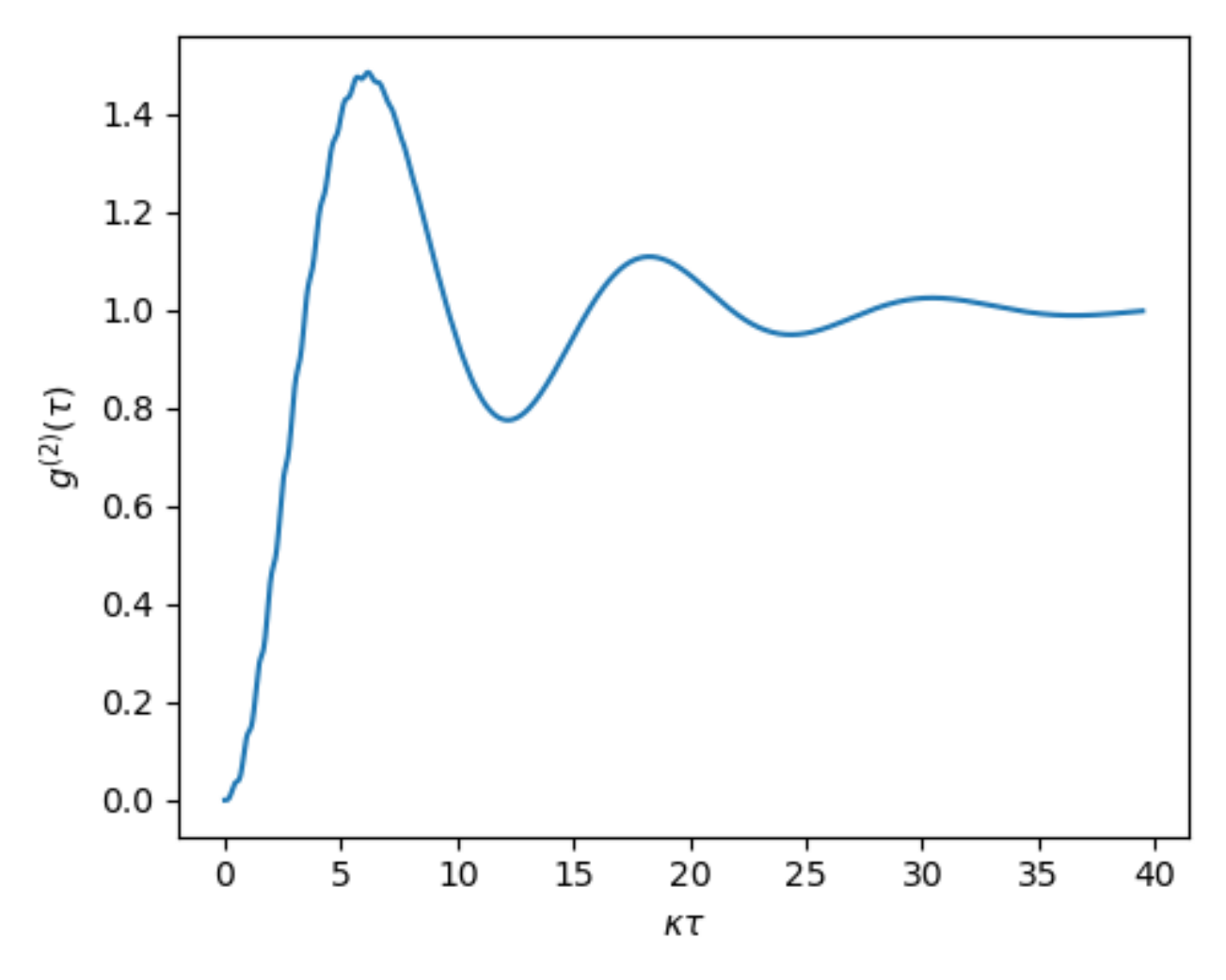}}
\subfigure[]{
\label{fig9.2}
\includegraphics[width=0.41\textwidth]{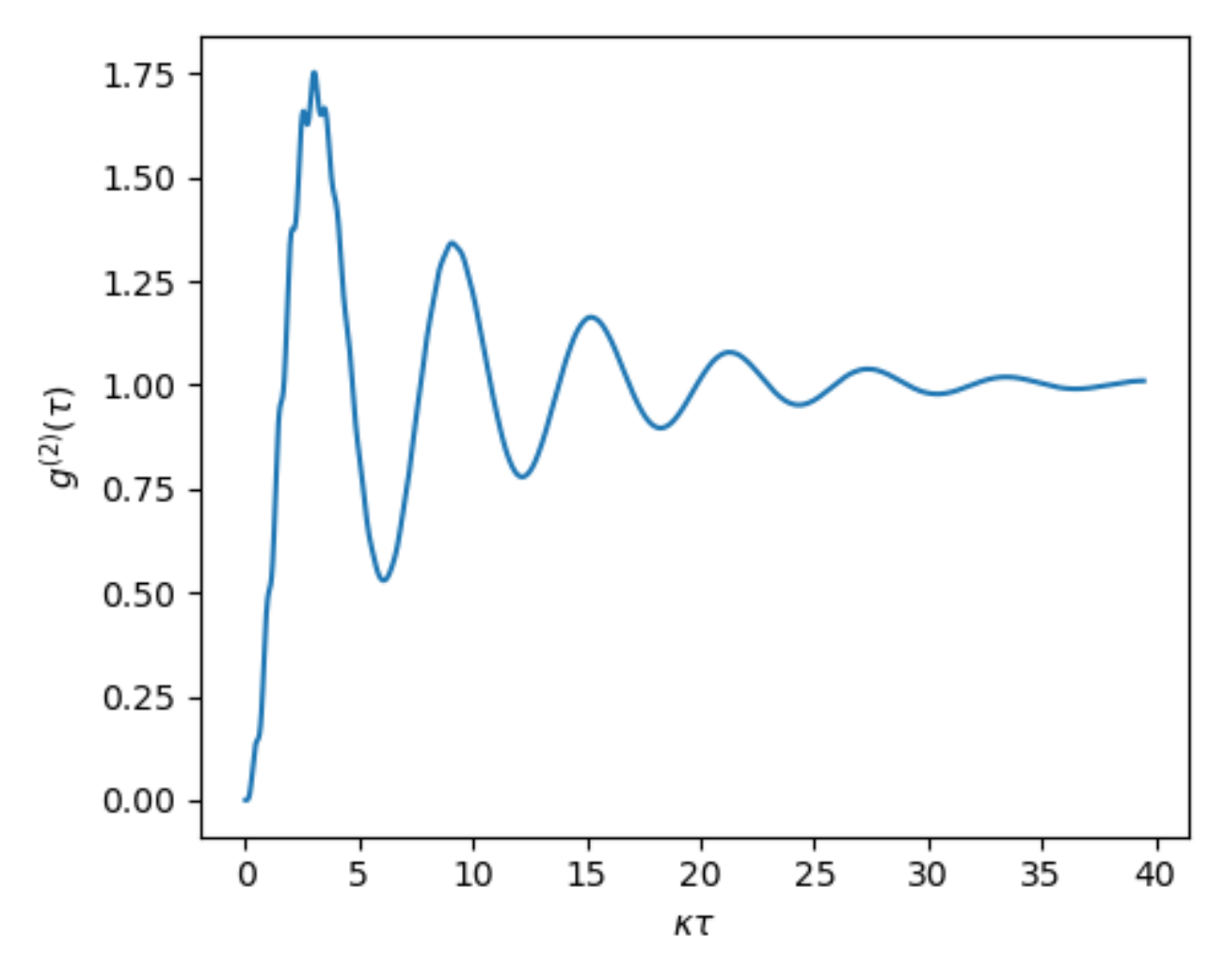}}
\caption{Plot of $g^{(2)}(\tau)$ as a function of $\kappa\tau$  under the optimal conditions in Eq. (\ref{e12})($\Delta_{c}=10\kappa$, $\Delta_{a}=1.95\kappa$, $V=13.9\kappa$, $g=2\pi\times7.8MHz$) for  different  pump laser
amplitude. (a)  $\varepsilon=0.2\kappa$, (b) $\varepsilon=0.4\kappa$.
 Other parameters are set as  $\gamma=2\pi\times0.4KHz$, $\kappa=2\pi\times2.5MHz$. }
\label{fig9}
\end{figure}

Finally, in order to fully characterize the quantum signatures  of the single photon emitter, we need to consider the  delayed second order correlation function in the steady state

\begin{equation}
g^{(2)}(\tau)=\frac{Tr[\hat{\rho} a^{\dag}(0)a^{\dag}(\tau)a(\tau)a(0)]}{(Tr[\hat{\rho} a^{\dag}(0)a(0)])^2}. \label{e100}
\end{equation}
In Figs. \ref{fig9.1} and  \ref{fig9.2}, we  plot the delayed second order correlation function  $g^{(2)}(\tau)$  for the UPB and  the PB optimal conditions. One can observe that at $\tau=0$, $g^{(2)}(0)$=0, and the delayed second order correlation function  $g^{(2)}(\tau) \textgreater g^{(2)}(0)$, i.e., fewer photon pairs are detected close together than further apart,  it's a  signature of the photon antibunching. We also find that $g^{(2)}(\tau)$ presents an oscillation behavior and the oscillation period is modulated by the pump laser amplitude $\varepsilon$.
In addition, the value of $g^{(2)}(\tau)$  approaches unity for long time delay.

\subsection{cavity-driven scheme}

\begin{figure}[!htbp]
\centering  

\includegraphics[width=0.6\textwidth]{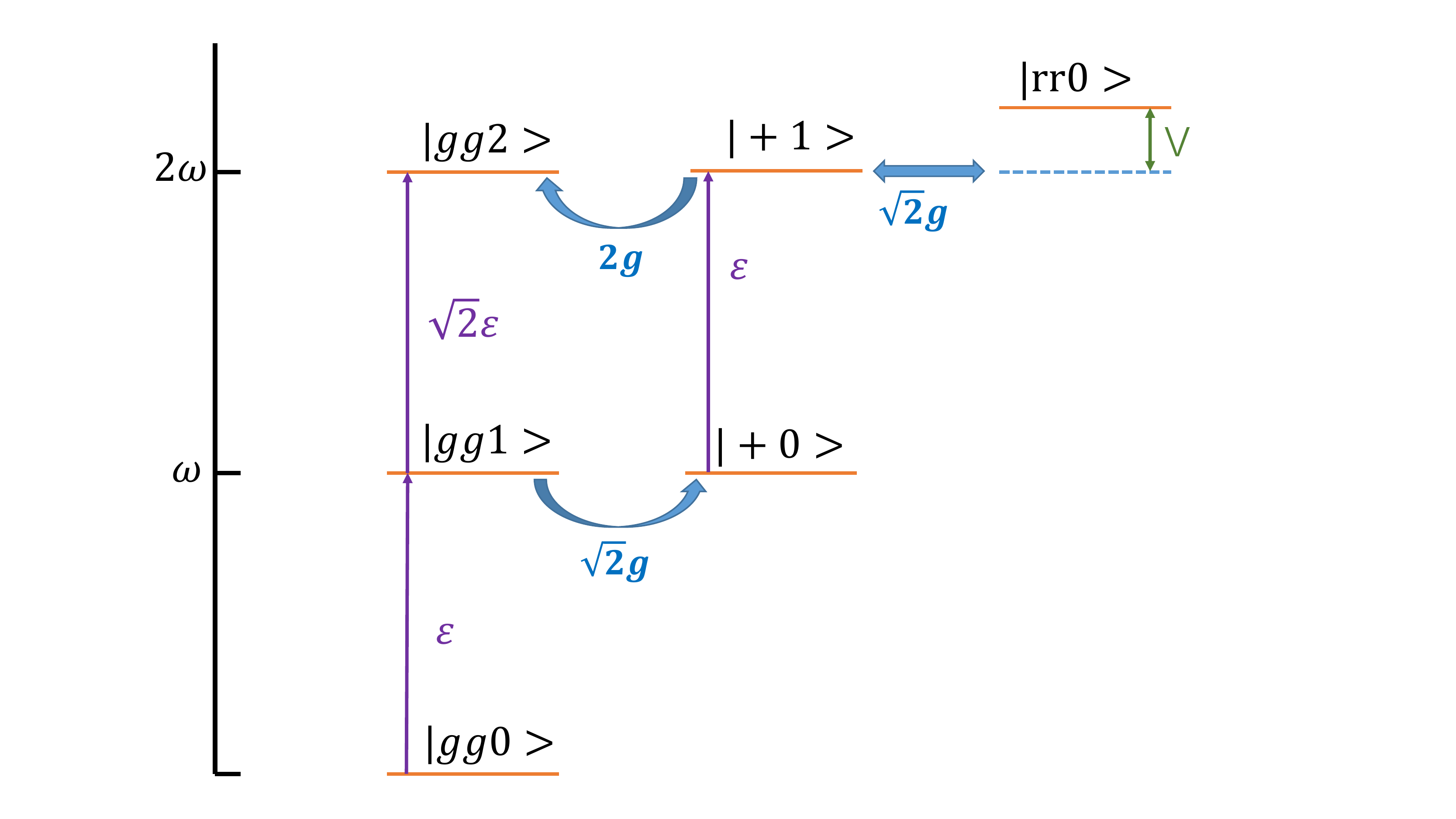}
\caption{ Transition pathways with cavity-driven scheme }
\label{fig7}
\end{figure}

Here, we  consider the cavity-driven scheme with  $\Delta_{a}=\Delta_{c}=0$, i.e., detunings  of the two level  Rydberg atoms transition frequency  and the cavity resonant frequency from the driving laser vanish. Analytic expression for the  time dependent coefficient can be obtained by solving the  Schr\"{o}dinger  equations:

\begin{equation}
\begin{split}
&i\dot{C}_{gg1}=-\frac{i\kappa}{2}C_{gg1}+\sqrt{2}gC_{+0}+\varepsilon C_{gg0}+\sqrt{2}\varepsilon C_{gg2},\\
&i\dot{C}_{gg2}=-i\kappa  C_{gg2}+2gC_{+1}+\sqrt{2}\varepsilon C_{gg1},\\
&i\dot{C}_{rr0}=(-i\gamma+V)C_{rr0}+\sqrt{2}gC_{+1},\\
&i\dot{C}_{+0}=-\frac{i\gamma}{2}C_{+0}+\sqrt{2}g C_{gg1}+\varepsilon C_{+1},\\
&i\dot{C}_{+1}=-\frac{i(\gamma+\kappa)}{2}C_{+1}+\sqrt{2}gC_{rr0}+2gC_{gg2}+\varepsilon C_{+0}. \label{e15}
\end{split}
\end{equation}
The steady state  solutions of Eq. (\ref{e15}) can be obtained  as follows
\begin{equation}
C_{gg2}=\frac{2 \sqrt{2}\varepsilon^{2}[(4\varepsilon^{2}+\gamma(\gamma+\kappa))(\gamma+iV)-4g^2(\gamma+2iV)]}{N}. \label{e16}
\end{equation}
where $N=(4\varepsilon^{2}+\kappa^2)(4\varepsilon^{2}+\gamma(\gamma+\kappa))(\gamma+iV)+32g^4(\kappa+2\gamma+2iV)+4g^2(\gamma\kappa(3\kappa+4\gamma)-
4\varepsilon^{2}(3\gamma+4iV)+2i\kappa(\kappa+2\gamma)V)$.
Within the same methods  mentioned  above, one  gets  the equations


\begin{equation}
\begin{split}
&4\varepsilon^{2}-4g^{2}+\kappa\gamma+\gamma^{2}=0,\\
&V=0. \label{e18}
\end{split}
\end{equation}

Here, Eq. (\ref{e18}) is the optimal condition of the UPB induced by the destructive interference,  which means that in order to get a strong photon antibunching effect, the  strength of the Rydberg-Rydberg interaction must always be zero. That is, a finite Rydberg coupling  strength $V$ could weaken the photon antibunching.

\begin{figure}[!htbp]
\centering  
\subfigure[]{
\label{fig8.1}
\includegraphics[width=0.43\textwidth]{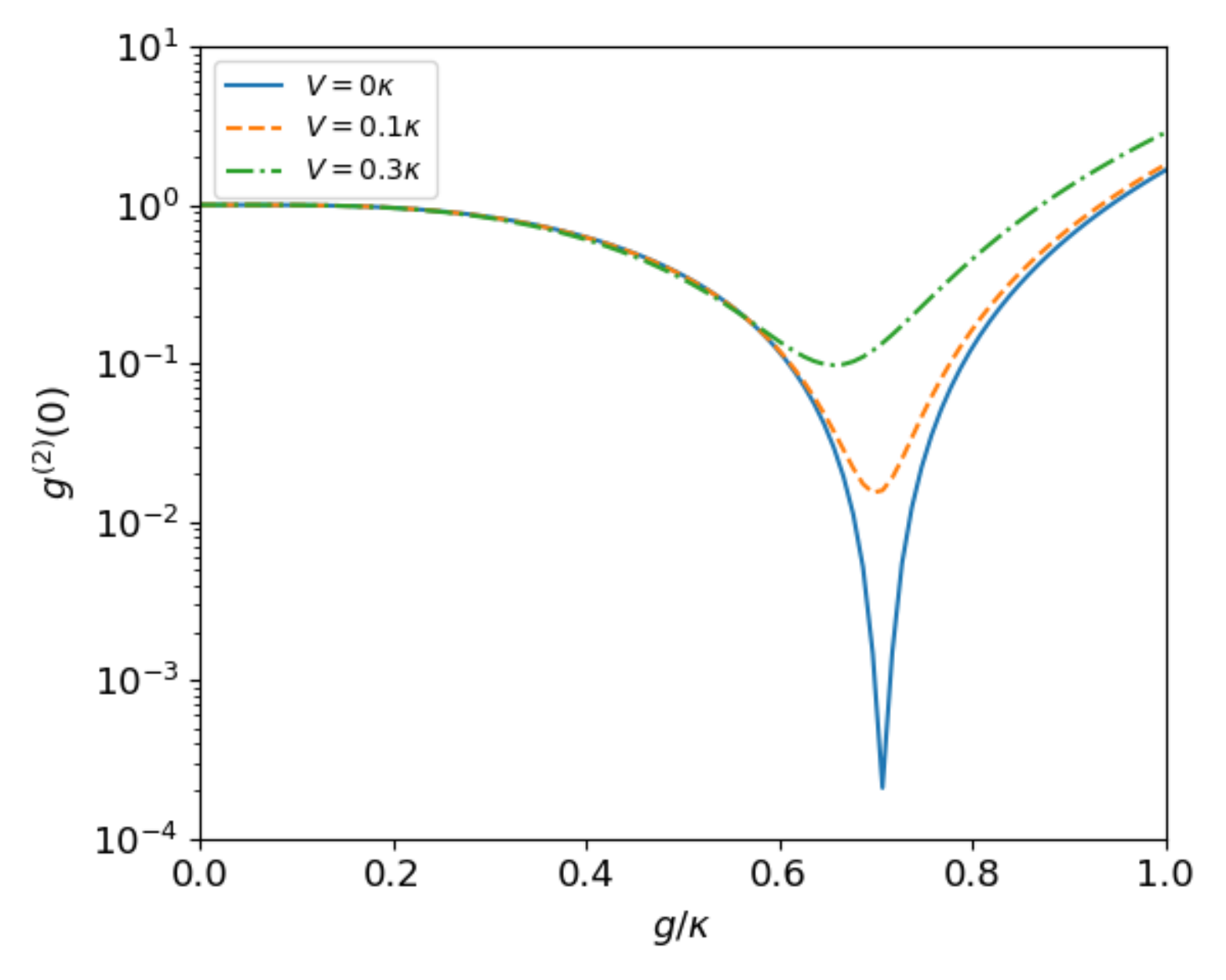}}
\subfigure[]{
\label{fig8.2}
\includegraphics[width=0.45\textwidth]{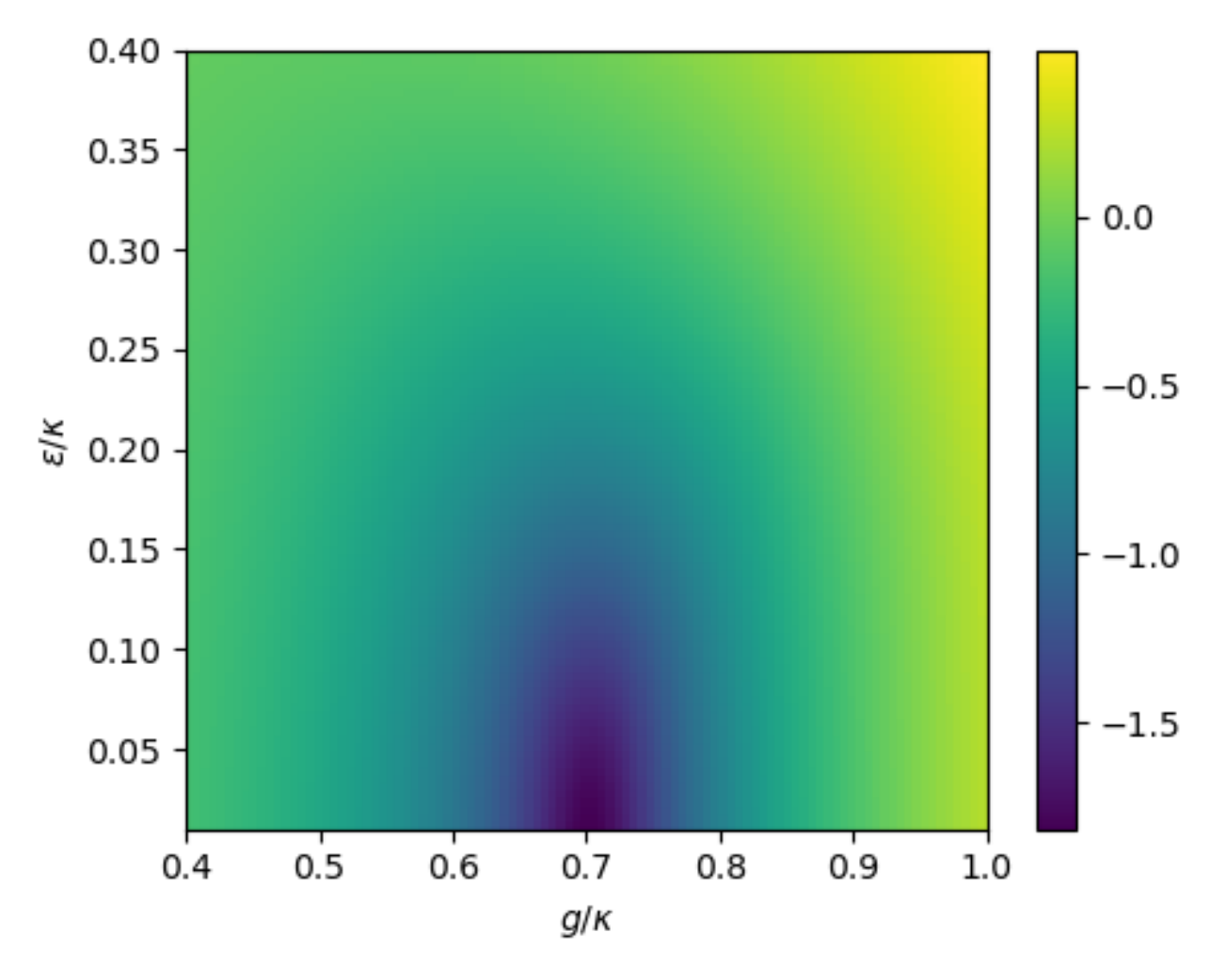}}
\caption{(a) Plot of $g^{(2)}(0)$ as a function of $g/\kappa$ with $V=0$, $V=0.1\kappa$ and $V=0.3\kappa$, respectively. Other parameters are given by  $\varepsilon=0.01\kappa$ and  $\gamma=\kappa$.  (b) Plot of $log_{10}[g^{(2)}(0)]$ as a function of $g/\kappa$ and $\varepsilon/\kappa$, Other parameters are given by  $V=0.1\kappa$ and  $\gamma=\kappa$.  }
\label{fig8}
\end{figure}

In what follows, we will numerically study the UPB and compare it with the analytic results, which are shown in
Fig. \ref{fig8.1}. Where  the  equal time second order correlation function $g^{(2)}(0)$ as a function of the cavity-atom coupling strength $g$  is plotted for different Rydberg coupling strength.
In Fig. \ref{fig7},  two excitation pathways are  $|gg0\rangle\stackrel{\varepsilon}{\longrightarrow}|gg1\rangle\stackrel{\sqrt{2}\varepsilon}{\longrightarrow}|gg2\rangle$  and $|gg0\rangle\stackrel{\varepsilon}{\longrightarrow}|gg1\rangle\stackrel{\sqrt{2}g}{\longleftrightarrow}|+ 0\rangle\stackrel{\varepsilon}{\longrightarrow}|+ 1\rangle(\stackrel{\sqrt{2}g}{\longleftrightarrow}|rr0\rangle\stackrel{\sqrt{2}g}{\longleftrightarrow}|+1\rangle)\stackrel{2g}{\longleftrightarrow}|gg2\rangle$. The destructive interference  results to  a  strong photon antibunching  when the  cavity-atom coupling strength $g\simeq $0.7$\kappa$ and $V=0$.
When  $V$ has a  deviation  from the optimal condition, the strength of the photon antibunching   decreases rapidly as shown in Fig. \ref{fig8.1}.
We also show the influence of cavity-driven strength on the UPB in Fig. \ref{fig8.2}.  Where  the  equal time second order correlation function $g^{(2)}(0)$ as a function of the cavity-driven strength $\varepsilon/\kappa$ and the cavity-atom coupling strength $g/\kappa$. It can be seen that one can obtain  strong photon antibunching effect with a weak cavity-driven strength.  However, This would also lead to a small value of  the  mean photon number in the cavity, and hence of single photons number emitted from the cavity.

\section{Conclusion}

In summary, we have investigated the photon antibunching  in a  Rydberg atoms-cavity system  under the atom-driven and cavity-driven scheme, respectively. By solving the  Schr\"{o}dinger equation and the Lindblad master equation, it can be found that in the case of the atom-driven, one can obtain  a strong UPB effect when the laser-cavity frequency detuning is large enough.  In the negative laser-cavity frequency detuning interval, the increase of Rydberg coupling strength  weakens the UPB effect.
In addition, Rydberg-Rydberg interaction can also result in strong PB effect  in the positive laser-cavity frequency  detuning region.
It's worth noting that within the range of the appropriate parameters,  while we get the extremely strong photon antibunching effect, the  mean photon number in the cavity has also improved significantly.
In the cavity-driven case, the increasing  interatomic Rydberg-Rydberg  interaction  strength  could weaken the  photon antibunching  rapidly. Our study  may provide   guidance for the experimental construction of  single-photon sources  in the Rydberg atoms-cavity  system.

\bibliographystyle{unsrt} 
\bibliography{ref}

\begin{thebibliography}{10}

\bibitem{1982Photon-antibunching}
H.~Paul.
\newblock Photon antibunching.
\newblock {\em Rev. Mod. Phys.}, 54:1061--1102, Oct 1982.

\bibitem{1977photon-antibunching}
H.~J. Kimble, M.~Dagenais, and L.~Mandel.
\newblock Photon antibunching in resonance fluorescence.
\newblock {\em Phys. Rev. Lett.}, 39:691--695, Sep 1977.

\bibitem{199Photon-antibunching}
X.~T. Zou and L.~Mandel.
\newblock Photon-antibunching and sub-poissonian photon statistics.
\newblock {\em Phys. Rev. A}, 41:475--476, Jan 1990.

\bibitem{1986Generating}
D.~Stoler and B.~Yurke.
\newblock Generating antibunched light from the output of a nondegenerate
  frequency converter.
\newblock {\em Phys. Rev. A}, 34:3143--3147, Oct 1986.

\bibitem{1986Bunching}
M.~H. Mahran and M.~Venkata Satyanarayana.
\newblock Bunching and antibunching properties of various coherent states of
  the radiation field.
\newblock {\em Phys. Rev. A}, 34:640--643, Jul 1986.

\bibitem{2004Quantum-dot}
A.~Kiraz, M.~Atat\"ure, and A.~Imamo\ifmmode~\breve{g}\else \u{g}\fi{}lu.
\newblock Quantum-dot single-photon sources: Prospects for applications in
  linear optics quantum-information processing.
\newblock {\em Phys. Rev. A}, 69:032305, Mar 2004.

\bibitem{2007Linear}
Pieter Kok, W.~J. Munro, Kae Nemoto, T.~C. Ralph, Jonathan~P. Dowling, and
  G.~J. Milburn.
\newblock Linear optical quantum computing with photonic qubits.
\newblock {\em Rev. Mod. Phys.}, 79:135--174, Jan 2007.

\bibitem{2005Single}
Brahim Lounis and Michel Orrit.
\newblock Single-photon sources.
\newblock {\em Reports on Progress in Physics}, 68(5):1129--1179, apr 2005.

\bibitem{2010On-chip}
P.~Yao, V.S.C. Manga~Rao, and S.~Hughes.
\newblock On-chip single photon sources using planar photonic crystals and
  single quantum dots.
\newblock {\em Laser \& Photonics Reviews}, 4(4):499--516, 2010.

\bibitem{1994Possibility}
W.~Leo\ifmmode~\acute{n}\else \'{n}\fi{}ski and R.~Tana\ifmmode~\acute{s}\else
  \'{s}\fi{}.
\newblock Possibility of producing the one-photon state in a kicked cavity with
  a nonlinear kerr medium.
\newblock {\em Phys. Rev. A}, 49:R20--R23, Jan 1994.

\bibitem{2002Deterministic}
Axel Kuhn, Markus Hennrich, and Gerhard Rempe.
\newblock Deterministic single-photon source for distributed quantum
  networking.
\newblock {\em Phys. Rev. Lett.}, 89:067901, Jul 2002.

\bibitem{2016Fate}
Alexandre Le~Boit\'e, Myung-Joong Hwang, Hyunchul Nha, and Martin~B. Plenio.
\newblock Fate of photon blockade in the deep strong-coupling regime.
\newblock {\em Phys. Rev. A}, 94:033827, Sep 2016.

\bibitem{2014From}
Yu-xi Liu, Xun-Wei Xu, Adam Miranowicz, and Franco Nori.
\newblock From blockade to transparency: Controllable photon transmission
  through a circuit-qed system.
\newblock {\em Phys. Rev. A}, 89:043818, Apr 2014.

\bibitem{2011Ph}
A.~J. Hoffman, S.~J. Srinivasan, S.~Schmidt, L.~Spietz, J.~Aumentado, H.~E.
  T\"ureci, and A.~A. Houck.
\newblock Dispersive photon blockade in a superconducting circuit.
\newblock {\em Phys. Rev. Lett.}, 107:053602, Jul 2011.

\bibitem{2013Photon-induced}
Xun-Wei Xu, Yuan-Jie Li, and Yu-xi Liu.
\newblock Photon-induced tunneling in optomechanical systems.
\newblock {\em Phys. Rev. A}, 87:025803, Feb 2013.

\bibitem{2013Photon-blockade}
Jie-Qiao Liao and Franco Nori.
\newblock Photon blockade in quadratically coupled optomechanical systems.
\newblock {\em Phys. Rev. A}, 88:023853, Aug 2013.

\bibitem{2011Phot}
P.~Rabl.
\newblock Photon blockade effect in optomechanical systems.
\newblock {\em Phys. Rev. Lett.}, 107:063601, Aug 2011.

\bibitem{2010Single}
T.~C.~H. Liew and V.~Savona.
\newblock Single photons from coupled quantum modes.
\newblock {\em Phys. Rev. Lett.}, 104:183601, May 2010.

\bibitem{2017Unconventional}
H.~Flayac and V.~Savona.
\newblock Unconventional photon blockade.
\newblock {\em Phys. Rev. A}, 96:053810, Nov 2017.

\bibitem{2015Unconventional}
Y.~H. Zhou, H.~Z. Shen, and X.~X. Yi.
\newblock Unconventional photon blockade with second-order nonlinearity.
\newblock {\em Phys. Rev. A}, 92:023838, Aug 2015.

\bibitem{2011Origin}
Motoaki Bamba, Atac Imamo\ifmmode~\breve{g}\else \u{g}\fi{}lu, Iacopo
  Carusotto, and Cristiano Ciuti.
\newblock Origin of strong photon antibunching in weakly nonlinear photonic
  molecules.
\newblock {\em Phys. Rev. A}, 83:021802, Feb 2011.

\bibitem{2019A-Photon}
Ming-Cui Li and Ai-Xi Chen.
\newblock A photon blockade in a coupled cavity system mediated by an atom.
\newblock {\em Applied Sciences}, 9(5), 2019.

\bibitem{2019Antibunching}
Xinyun Liang, Zhenglu Duan, Qin Guo, Cunjin Liu, Shengguo Guan, and Yi~Ren.
\newblock Antibunching effect of photons in a two-level emitter-cavity system.
\newblock {\em Phys. Rev. A}, 100:063834, Dec 2019.

\bibitem{2018Observation1}
H.~J. Snijders, J.~A. Frey, J.~Norman, H.~Flayac, V.~Savona, A.~C. Gossard,
  J.~E. Bowers, M.~P. van Exter, D.~Bouwmeester, and W.~L\"offler.
\newblock Observation of the unconventional photon blockade.
\newblock {\em Phys. Rev. Lett.}, 121:043601, Jul 2018.

\bibitem{2018Observation2}
Cyril Vaneph, Alexis Morvan, Gianluca Aiello, Mathieu F\'echant, Marco Aprili,
  Julien Gabelli, and J\'er\^ome Est\`eve.
\newblock Observation of the unconventional photon blockade in the microwave
  domain.
\newblock {\em Phys. Rev. Lett.}, 121:043602, Jul 2018.

\bibitem{2000Intensity}
G.~T. Foster, S.~L. Mielke, and L.~A. Orozco.
\newblock Intensity correlations in cavity qed.
\newblock {\em Phys. Rev. A}, 61:053821, Apr 2000.

\bibitem{2014Unconventional-photon}
Dario Gerace and Vincenzo Savona.
\newblock Unconventional photon blockade in doubly resonant microcavities with
  second-order nonlinearity.
\newblock {\em Phys. Rev. A}, 89:031803, Mar 2014.

\bibitem{2013Optimal}
S~Ferretti, V~Savona, and D~Gerace.
\newblock Optimal antibunching in passive photonic devices based on coupled
  nonlinear resonators.
\newblock {\em New Journal of Physics}, 15(2):025012, feb 2013.

\bibitem{2014Tun}
Xun-Wei Xu and Yong Li.
\newblock Tunable photon statistics in weakly nonlinear photonic molecules.
\newblock {\em Phys. Rev. A}, 90:043822, Oct 2014.

\bibitem{2017Quantum-interference-assisted}
Bijita Sarma and Amarendra~K. Sarma.
\newblock Quantum-interference-assisted photon blockade in a cavity via
  parametric interactions.
\newblock {\em Phys. Rev. A}, 96:053827, Nov 2017.

\bibitem{2015Tunable}
Hui Wang, Xiu Gu, Yu-xi Liu, Adam Miranowicz, and Franco Nori.
\newblock Tunable photon blockade in a hybrid system consisting of an
  optomechanical device coupled to a two-level system.
\newblock {\em Phys. Rev. A}, 92:033806, Sep 2015.

\bibitem{2020Photon}
Dong-Yang Wang, Cheng-Hua Bai, Shutian Liu, Shou Zhang, and Hong-Fu Wang.
\newblock Photon blockade in a double-cavity optomechanical system with
  nonreciprocal coupling.
\newblock {\em New Journal of Physics}, 22(9):093006, sep 2020.

\bibitem{2014Antibunching}
Marc-Antoine Lemonde, Nicolas Didier, and Aashish~A. Clerk.
\newblock Antibunching and unconventional photon blockade with gaussian
  squeezed states.
\newblock {\em Phys. Rev. A}, 90:063824, Dec 2014.

\bibitem{2010Quantuminformation}
M.~Saffman, T.~G. Walker, and K.~M\o{}lmer.
\newblock Quantum information with rydberg atoms.
\newblock {\em Rev. Mod. Phys.}, 82:2313--2363, Aug 2010.

\bibitem{2016Quantumcomputing}
M~Saffman.
\newblock Quantum computing with atomic qubits and rydberg interactions:
  progress and challenges.
\newblock {\em Journal of Physics B: Atomic, Molecular and Optical Physics},
  49(20):202001, oct 2016.

\bibitem{2016Nonlinear}
O~Firstenberg, C~S Adams, and S~Hofferberth.
\newblock Nonlinear quantum optics mediated by rydberg interactions.
\newblock {\em Journal of Physics B: Atomic, Molecular and Optical Physics},
  49(15):152003, jun 2016.

\bibitem{2013PhotonRydberg}
Jin-Feng Huang, Jie-Qiao Liao, and C.~P. Sun.
\newblock Photon blockade induced by atoms with rydberg coupling.
\newblock {\em Phys. Rev. A}, 87:023822, Feb 2013.

\bibitem{2014Quantumstatistics}
A~Grankin, E~Brion, E~Bimbard, R~Boddeda, I~Usmani, A~Ourjoumtsev, and
  P~Grangier.
\newblock Quantum statistics of light transmitted through an intracavity
  rydberg medium.
\newblock {\em New Journal of Physics}, 16(4):043020, apr 2014.

\bibitem{2012CorrelatedPhoton}
J.~D. Pritchard, C.~S. Adams, and K.~M\o{}lmer.
\newblock Correlated photon emission from multiatom rydberg dark states.
\newblock {\em Phys. Rev. Lett.}, 108:043601, Jan 2012.

\bibitem{2019Interfering}
K.~Hou, C.~J. Zhu, Y.~P. Yang, and G.~S. Agarwal.
\newblock Interfering pathways for photon blockade in cavity qed with one and
  two qubits.
\newblock {\em Phys. Rev. A}, 100:063817, Dec 2019.

\bibitem{2019Hybrid}
C.~J. Zhu, K.~Hou, Y.~P. Yang, and L.~Deng.
\newblock Hybrid level anharmonicity and interference-induced photon blockade
  in a two-qubit cavity qed system with dipole-dipole interaction.
\newblock {\em Photon. Res.}, 9(7):1264--1271, Jul 2021.

\bibitem{2017Hyperradiance}
Marc-Oliver Pleinert, Joachim von Zanthier, and Girish~S. Agarwal.
\newblock Hyperradiance from collective behavior of coherently driven atoms.
\newblock {\em Optica}, 4(7):779--785, Jul 2017.

\bibitem{2017Collective}
C.~J. Zhu, Y.~P. Yang, and G.~S. Agarwal.
\newblock Collective multiphoton blockade in cavity quantum electrodynamics.
\newblock {\em Phys. Rev. A}, 95:063842, Jun 2017.

\bibitem{2019Deterministic}
Bastian Hacker, Stephan Welte, Severin Daiss, Armin Shaukat, Stephan Ritter,
  Lin Li, and Gerhard Rempe.
\newblock Deterministic creation of entangled atom–light schrödinger-cat
  states.
\newblock {\em Nat. Photonics}, 13:110--115, Jan 2019.

\bibitem{2013Direct}
L.~B\'eguin, A.~Vernier, R.~Chicireanu, T.~Lahaye, and A.~Browaeys.
\newblock Direct measurement of the van der waals interaction between two
  rydberg atoms.
\newblock {\em Phys. Rev. Lett.}, 110:263201, Jun 2013.

\bibitem{2009Quasi}
I.~I. Beterov, I.~I. Ryabtsev, D.~B. Tretyakov, and V.~M. Entin.
\newblock Quasiclassical calculations of blackbody-radiation-induced
  depopulation rates and effective lifetimes of rydberg $ns$, $np$, and $nd$
  alkali-metal atoms with $n\ensuremath{\le}80$.
\newblock {\em Phys. Rev. A}, 79:052504, May 2009.

\bibitem{2021Efficient}
Yu-Guo Liu, Keyu Xia, and Shi-Liang Zhu.
\newblock Efficient microwave-to-optical single-photon conversion with a single
  flying circular rydberg atom.
\newblock {\em Opt. Express}, 29(7):9942--9959, Mar 2021.

\end{thebibliography}

\end{document}